\documentclass[aps,pra,twocolumn,showpacs,floatfix]{revtex4}
\usepackage{graphicx}
\usepackage{nicefrac}
\usepackage{amsmath}
\usepackage{amsfonts}
\usepackage{amssymb}
\usepackage{amsthm}
\usepackage{epsf}
\usepackage{bm}
\usepackage{bbm}
\usepackage{longtable}

\usepackage{dcolumn}

\newcolumntype{.}{D{x}{}{-1}}

\newcommand{\vare}{\varepsilon}

\newcommand{\Za}{{Z \alpha}}

\def\rms{<\!\!\!\,r^2\!\!\!>^{1/2}}

\begin{document}

\title{Hyperfine structure of $\bm{S}$ states in Li and Be$^{\bm{+}}$}

\author{V.~A.~Yerokhin}
 \affiliation{Center for Advanced Studies, St.~Petersburg State Polytechnical
University, Polytekhnicheskaya 29, St.~Petersburg 195251, Russia}

\begin{abstract}

A large-scale configuration-interaction (CI) calculation is reported for the hyperfine
splitting of the $2\,^2S$ and $3\,^2S$ states of $^7$Li and $^9$Be$^+$. The CI
calculation based on the Dirac-Coulomb-Breit Hamiltonian is supplemented with a separate
treatment of the QED, nuclear-size, nuclear-magnetization distribution, and recoil
corrections. The nonrelativistic limit of the CI results is in excellent agreement with
variational calculations. The theoretical values obtained for the hyperfine splitting
are complete to the relative order of $\alpha^2$ and improve upon results of previous
studies.

\end{abstract}


\maketitle
%

The hyperfine structure (hfs) of few-electron atoms has been an attractive subject
of theoretical studies for decades, one of the reasons being a few ppm accuracy
achieved in experiments on Li and Be$^+$ \cite{beckmann:74,wineland:83:prl}.
Interest in this topic was enhanced even further recently, in view of prospects of
using hfs data to get an access to the neutron halo structure, the proton charge
distribution, and the nuclear vector polarizability, particularly for isotopes of
Be$^+$ \cite{nakamura:06,pachucki:07:BW}. Despite the considerable attention
received, a high-precision theoretical description of hfs in few-electron atoms
remains a difficult task. The main problem lies in the high singularity of the hfs
interaction and, as a consequence, in the dependence of the calculated results on
the quality of the correlated wave function near the nucleus.

Among numerous theoretical investigations  performed previously for Li and Be$^+$,
two apparently most accurate ones are the multiconfigurational Dirac-Fock (MCDF)
calculation \cite{bieron:96} and the Hylleraas-type variational calculation
\cite{yan:96}. Both studies report good agreement with the experiment, but they are
not entirely consistent with each other in treatment of individual corrections. The
MCDF calculation does not include the binding QED effects and, in the case of Li,
the nuclear magnetization distribution effect. The variational calculation yields
accurate results for the nonrelativistic Fermi contact term but treats the
relativistic effects in an effective way only, by rescaling the hydrogenic result.
This indicates that neither of these studies is complete at the relative order of
$\alpha^2$ ($\alpha$ is the fine-structure constant). The aim of the present
investigation is to perform a high-precision calculation of the hfs splitting in Li
and Be$^+$, with a complete treatment of all corrections $\sim \!\alpha^2$.

A possible way to accomplish this task would be to supplement the nonrelativistic
calculation \cite{yan:96} with a rigorous evaluation of the relativistic correction,
whose expression was recently derived by Pachucki \cite{pachucki:02:pra}. Such a
calculation has not been performed so far. In the present work, the relativistic
correction will be accounted for by means of the Dirac-Coulomb-Breit Hamiltonian.

The magnetic dipole hfs splitting of an energy level of an $nS$ state is
conveniently represented in terms of a dimensionless function $G_n(Z)$ defined as
\cite{shabaev:94:hfs}
\begin{equation} \label{eq2}
    \Delta E_n = \frac43\,\frac{\alpha(\Za)^3}{n^3}\,
        \frac{m}{m_p}\,\frac{\mu}{\mu_N}\,
      \frac{2I+1}{2I}\,\frac{mc^2}{(1+m/M)^{3}}\, G_n(Z)
      \,,
\end{equation}
where $\mu$ is the nuclear magnetic moment, $\mu_N = |e|/(2\,m_p)$ is the nuclear
magneton; $m$, $m_p$, and $M$ are the masses of the electron, the proton, and the
nucleus, respectively; $I$ is the nuclear spin quantum number, and $Z$ is the
nuclear charge number. The function $G$ defined in this way is unity for a
non-relativistic point-nucleus H-like atom.

Within the leading relativistic approximation, the electron correlation can be
described by the Dirac-Coulomb-Breit equation, which is solved by the
configuration-interaction (CI) Dirac-Fock (DF) method in the present work. The
many-electron wave function $\Psi(PJM)$ with the parity $P$, the momentum quantum
number $J$, and the momentum projection $M$ is represented as a sum of
configuration-state functions (CSFs),
\begin{equation}\label{eq1}
  \Psi(PJM) = \sum_r c_r \Phi(\gamma_r PJM)\,.
\end{equation}
The CSFs are obtained as linear combinations of the Slater determinants constructed
from the positive-energy solutions of the Dirac equation with the frozen-core DF
potential. The mixing coefficients $c_r$ are determined by diagonalizing the
Hamiltonian matrix. The hfs splitting is obtained as the expectation value of the
hfs operator on the many-electron wave function (\ref{eq1}). The corresponding
formulas are well-known, see, {\em e.g.}, \cite{bieron:96}. To perform a CI
calculation, we devised a code, incorporating and adapting a number of existing
packages \cite{CI} for setting up the CSFs, calculating angular-momentum
coefficients, and diagonalizing the Hamiltonian matrix. The largest number of CSFs
simultaneously handled was about a half of a million, with the number of nonzero
elements in the Hamiltonian matrix of about 5 billions. A thorough optimization of
the code was carried out, in order to keep the time and memory consumption of the
calculation within reasonable limits.

The dominant part of the hfs splitting in light atoms is delivered by the
Dirac-Coulomb Hamiltonian. This was the most demanding part of the calculation since
a high relative precision was required. The one-electron orbitals for constructing
CFSs were obtained by the dual-kinetic-balance (DKB) B-spline basis set method
\cite{shabaev:04:DKB} for the Dirac equation. For a given number of B-splines $n_a$,
all eigenstates were taken with the energy $0< \vare \le mc^2(1+Z\alpha\, E_{\rm
max})$ and the orbital quantum number $l \le l_{\rm max}$, where $E_{\rm max}$ was
varied between $0.5$ and $6$ and $l_{\rm max}$, between $1$ and $7$. Three main sets
of one-electron orbitals were employed in the present work: (A)
$20s\,$$20p\,$$19d\,$$19f\,$$18g\,$$18h$ with $n_a=44$ and $E_{\rm max}=3$, (B)
$14s\,$$14p\,$$14d\,$$13f\,$$13g\,$$13h\,$$12i\,$$12k$ with $n_a=34$ and $E_{\rm
max}=0.5$,  and (C) $25s\,$$25p\,$$24d\,$ with $n_a=54$ and $E_{\rm max}=6$. Here,
the notation, {\em e.g.}, $20p$ means $20p_{1/2}\,$$20p_{3/2}$. Calculational
results were first obtained with the set (A) and then corrected for contributions of
the higher partial waves with the set (B) and for a more complete representation of
the Dirac spectrum with the set (C). The set of CSFs used in the calculation was
obtained by taking all single, double, and triple excitations from the reference
configuration with at least one electron orbital with $l\le 1$ present. The triple
excitations that were left out in this way were found to yield a negligible
contribution. Inclusion of the {\em Breit interaction} into the Dirac-Coulomb
Hamiltonian yields only a small correction in the case of Li and Be$^+$. Since the
effect is small, it is sufficient to use a much shorter basis set for its
evaluation, which simplifies the computation greatly.

%
%
\begin{table}[t]
\caption{The Dirac-Coulomb-Breit part of the hfs splitting, in terms of $G(Z)$.
 \label{tab:pwexp} }
\begin{ruledtabular}
\begin{tabular}{lc..}
         & $l_{\rm max}$ &  \multicolumn{1}{c}{Li $\,2^2S \ \ \ $}
                                         &  \multicolumn{1}{c}{Be$^+$ $\,2^2S \ \ \ $}\\
\hline\\[-9pt]
Coulomb &     1 &  0.214\,4x70\,3       &  0.390\,1x59\,9       \\
        &     2 &  0.215\,1x67\,8       &  0.390\,7x98\,6       \\
        &     3 &  0.215\,3x04\,4       &  0.390\,9x38\,7       \\
        &     4 &  0.215\,3x46\,2       &  0.390\,9x84\,7       \\
        &     5 &  0.215\,3x62\,9       &  0.391\,0x03\,8       \\
        &     6 &  0.215\,3x71\,9       &  0.391\,0x14\,5       \\
        &     7 &  0.215\,3x76\,5       &  0.391\,0x20\,2       \\
&       $\infty$&  0.215\,3x84\,8(49)   &  0.391\,0x30\,4(61) \\
Breit   &       &  0.000\,0x15\,9       &  0.000\,0x38\,6       \\
Total   &       &  0.215\,4x00\,7(49)   &  0.391\,0x69\,0(61) \\
MCDF \cite{bieron:96} &
                &  0.215\,2x87          &  0.390\,9x84        \\
Hylleraas$^a$ \cite{yan:96} &
                 &  0.215\,3x79(13)     &  0.391\,0x23(34)  \\
\end{tabular}
\end{ruledtabular}
$^a$ the sum of the nonrelativistic, the relativistic, and the nuclear-charge
distribution terms.
\end{table}

The results of our CI calculation of the Dirac-Coulomb-Breit part of the ground-state
hfs in $^7$Li and $^9$Be$^+$ are presented in Table~\ref{tab:pwexp}. The Fermi model was
employed for the nuclear-charge distribution, with the nuclear-charge radii
\cite{angeli:04} $\rms = 2.431(28)$~fm for Li and $\rms = 2.518(11)$~fm for Be. The
uncertainties specified in the table include the estimated error due to the
incompleteness of the basis and due to the finite nuclear size. The error of the Breit
part was found to be negligible. Our results are in reasonable agreement with the
nonrelativistic variational results \cite{yan:96} but deviate significantly from the
MCDF values \cite{bieron:96}. The comparison leads us to a conclusion that the dominant
part of the relativistic correction can indeed be accounted for by an effective scaling
of the hydrogenic results, as was argued in \cite{yan:96}. A complete evaluation of the
relativistic correction within the $\Za$-expansion approach, however, has to be
performed along the way paved in \cite{pachucki:02:pra}, which has not been done yet.

Comparison with the results of \cite{yan:96} would become possible on a much higher
level of accuracy if we identified the nonrelativistic part of our CI results. Such
an identification was carried out by repeating our CI calculations for different
values of $\alpha$ (namely, three values with ratios $\alpha^{\prime}/\alpha = 0.9$,
$1$, and $1.1$ were used). For each value of $\alpha$, the finite nuclear-charge
distribution (FNC) correction was evaluated separately and subtracted from the CI
values. The point-nucleus results thus obtained were fitted to a polynomial in
$\alpha$, assuming the absence of the linear term. In this way, the CI results with
the physical value of $\alpha$ were separated into three parts: the nonrelativistic
point-nucleus contribution, the relativistic point-nucleus correction, and the FNC
correction. The numerical results for them are listed in Table~\ref{tab:results}.

%
%
\begin{table*}
\setlength{\LTcapwidth}{\textwidth}
 \caption{Individual contributions to the hfs
splitting, in terms of $G(Z)$. The experimental values for the function $G$ for Li
were inferred from the original references by using the nuclear magnetic moment
$\mu/\mu_N = 3.256\,426\,8(17)$ \cite{stone:05}.
 \label{tab:results} }
\begin{ruledtabular}
\begin{tabular}{l....}
                & \multicolumn{1}{c}{$^7$Li $2\,^2S \ \ \ $}
                               & \multicolumn{1}{c}{$^7$Li $3\,^2S\ \ \ $}
                                                  & \multicolumn{1}{c}{$^9$Be$^+$ $2\,^2S\ \ \ $}
                                                        & \multicolumn{1}{c}{$^9$Be$^+$ $3\,^2S\ \ \ $}
                                               \\
\hline\\[-9pt]
Nonrelativistic       &  0.215\,2x51\,^a       &      0.168\,3x40\,^a    &   0.390\,5x44\,^a    &  0.335\,0x66\,^a  \\
Ref.~\cite{yan:96}    &  0.215\,2x54\,(4)      &      0.168\,3x51\,(13)  &   0.390\,5x49\,(9)   &                   \\
Relativistic          &  0.000\,2x05\,^a       &      0.000\,1x59\,^a    &   0.000\,6x64\,^a    &  0.000\,5x64\,^a  \\
Finite nuclear charge & -0.000\,0x55\,^a       &     -0.000\,0x43\,^a    &  -0.000\,1x39\,^a    & -0.000\,1x19\,^a  \\
Dirac-Coulomb-Breit   &  0.215\,4x01\,(5)      &      0.168\,4x56\,(9)   &   0.391\,0x69\,(6)   &  0.335\,5x10\,(9)\\
QED                   &  0.000\,1x82\,(4)      &      0.000\,1x43\,(4)   &   0.000\,2x89\,(12)  &  0.000\,2x50\,(12)\\
Bohr-Weisskopf        & -0.000\,0x24\,(3)      &     -0.000\,0x19\,(2)   &  -0.000\,0x62\,(17)  & -0.000\,0x53\,(14)\\
Specific mass shift   &  0.000\,0x02           &      0.000\,0x02        &   0.000\,0x02        &  0.000\,0x02      \\
Negative-continuum    &  0.000\,0x02           &      0.000\,0x02        &   0.000\,0x05        &  0.000\,0x05      \\
Total theory          &  0.215\,5x63\,(7)      &      0.168\,5x84\,(10)  &   0.391\,3x04\,(22)  &  0.335\,7x14\,(21)\\
Ref.~\cite{yan:96}    &  0.215\,5x4\,(2)       &      0.168\,5x8\,(2)    &   0.391\,2x7\,(4)    &                   \\
Experiment            &  0.215\,5x61\,1\,(1)^b &      0.168\,6x0\,(2)^c  &   0.391\,2x60\,(1)^{e*} &       \\
                      &                        &      0.171\,5x\,(4)^d   &   0.391\,2x40\,(6)^{e\dag}   & \\
\end{tabular}
\end{ruledtabular}
$^a$ These three entries are inferred from the corresponding Dirac-Coulomb-Breit
values; their sum is expected to be more accurate than each of the entries
separately; $^b$ Ref.~\cite{beckmann:74}; $^c$ Ref.~\cite{bushaw:03}; $^d$
Ref.~\cite{stevens:95}; $^{e*}$ Ref.~\cite{wineland:83:prl} with $\mu/\mu_N =
-1.177\,432(3)$ \cite{stone:05}; $^{e\dag}$~Ref.~\cite{wineland:83:prl} with
$\mu/\mu_N = -1.177\,49(2)$ \cite{stone:05}.
\end{table*}

The {\em FNC correction} was evaluated for both the hydrogenic wave functions and
the CI many-electron wave functions. In the latter case, a series of the CI
calculations with different values of the nuclear-charge radius $R$ was performed
and the FNC correction was extracted by a fit, using the analytical form of the $R$
dependence \cite{shabaev:94:hfs}. It was found that, with an accuracy of
$\sim\,$0.5\%, there was no screening effect on the relative value of this
correction.

The {\em QED effects} induce the largest correction to be added to the
Dirac-Coulomb-Breit hfs value. For $nS$ states of few-electron atoms, the QED
correction can be written in the same form as for hydrogen \cite{sapirstein:90:kin},
\begin{eqnarray}\label{eq3}
    \delta G_n(Z) &=& \frac{\alpha}{\pi}\,G^{\rm NR}_n(Z)\,\left\{ \frac12
      + \Za \,\pi \left(\ln 2-\frac52 \right)
  \right. \nonumber \\ && \left.
    \!\!\!\!\!\!\!\!\!\!\!\!\!\!\!\!\!\!\!\!\!\!\!\!
      + (\Za)^2 \left[ -\frac83 \ln^2(\Za)+a_{21} \ln(\Za)+a_{20} \right]
      \right\} ,
\end{eqnarray}
where $G_n^{\rm NR}$ is the nonrelativistic hfs value. The first three coefficients
in the $\Za$ expansion (\ref{eq3}) are the same as for hydrogen. The higher-order
terms $a_{21}$ and $a_{20}$ are different and not known at present. One can,
however, estimate them with their hydrogenic values
\cite{karshenboim:02:epjd,jentschura:06:hfs}: $a_{21}(2s) = -1.1675$, $a_{20}(2s) =
11.3522$, $a_{21}(3s) = -2.3754$, and $a_{20}(3s) = 9.7474$. A 100\% uncertainty is
ascribed to this approximation. Essentially the same treatment of the QED correction
was reported in \cite{yan:96}; the QED results of \cite{bieron:96} differ by
$\sim\,$40\% due to the neglect of the binding QED effects [i.e., the terms in
(\ref{eq3}) beyond the first one].

The {\em nuclear structure effects} have significant influence on hfs and should be
taken into account. Their rigorous description is a demanding problem. The way for
its solution was paved in recent studies \cite{friar:05,pachucki:07:BW}. Practical
realizations of this approach, however, are so far restricted by two- and
three-nucleon systems \cite{friar:05} and their extension for more complex nuclei
like $^7$Li and $^9$Be looks problematic.

The most widely used approach up to now is to account for the nuclear magnetization
distribution [the Bohr-Weisskopf (BW) effect] by means of the Zemach formula
\cite{zemach:56}, which is simple and apparently model independent. Such approach
ignores inelastic effects, which can yield a large contribution \cite{friar:05}, and
it is not clear what uncertainty should be ascribed to such results. In the present
study, we calculate the BW correction  within the single-particle (SP) nuclear model
\cite{bohr:50,shabaev:97:pra}, in which the nuclear magnetic moment is assumed to be
induced by the odd nucleon. This model is expected to be reasonably adequate for
$^7$Li since it reproduces well the observable nuclear magnetic moment basing on
just the free-nucleon $g$ factors, the difference being only 15\%. For $^9$Be, the
deviation is four times larger and the SP approach is expected to yield worse
results.

Within the SP model, the BW effect can be accounted for by adding a multiplicative
magnetization-distribution function to the standard point-dipole hfs interaction
\cite{shabaev:97:pra}. The distribution function is induced by the wave function of
the odd nucleon and is obtained by solving the Schr\"odinger equation with the
Woods-Saxon potential and an empirical spin-orbit interaction included. The
parameters of the potential were taken from \cite{elton:67}. The BW correction was
calculated for both the one-electron wave functions and the CI many-electron wave
functions. It was found that, with a very good accuracy ($< 0.5\%$), there was no
screening effect on the relative value of this correction. Our calculational results
are larger than the Zemach-formula values of \cite{yan:96} by $\sim\,$10\% in the
case of Li and by $\sim\,$30\% in the case of Be. The Zemach-formula result of
\cite{bieron:96} for Be is larger than the one of \cite{yan:96} by a factor of four,
which is due, we believe, to a misinterpretation of the Zemach formula in
\cite{bieron:96}. Our computational results for the BW correction are presented in
Table~\ref{tab:results}. The error bars specified were obtained as the difference of
the SP and the Zemach values and should be regarded as order-of-magnitude
estimations of the error. We checked that similar evaluations of the nuclear effect
on hfs in $^3$He$^+$ agree well with a much more elaborate calculation of
\cite{friar:05}.

The leading {\em recoil} contribution is given by the mass scaling factor
$(1+m/M)^{-3}$ included into the definition of the function $G$ in (\ref{eq2}). The
remaining correction (within the nonrelativistic approach) is due to the specific
mass shift (SMS) and is very small for the $S$ states. We calculate it by
introducing the SMS term $(m/M)\sum_{i<j}\bm{p}_i\cdot\bm{p}_j$ into the
Dirac-Coulomb Hamiltonian and taking the increment of the CI results with and
without SMS ($\bm{p}$ is the momentum operator). Our results agree with the
estimates obtained in \cite{yan:96} but are more accurate. For the $2\,^2S$ and
$3\,^2S$ states of Li, we obtain $\delta G = 2.0(2)\times 10^{-6}$ and $1.9(2)\times
10^{-6}$, respectively, which should be compared with $2(5)\times 10^{-6}$ and
$2(20)\times 10^{-6}$ from \cite{yan:96}, respectively.

The {\em negative-continuum contribution} might be of some importance in
calculations involving the operators that mix the upper and the lower components of
the Dirac wave function. The hfs operator is of this kind, so we have to obtain an
estimation for this correction. We calculate the negative-continuum contribution by
employing the many-body perturbation theory to the first order. The same
one-electron DKB basis set was used as in the CI calculations, with the only
difference that all negative-energy eigenstates were taken.

Our total theoretical values for the hfs splitting presented in
Table~\ref{tab:results} agree with the results by Yan {\em et al.}
\cite{yan:96} but are more accurate. The present theory agrees very
well with experiments on Li, the only exception being the
experimental result \cite{stevens:95}, which contradicts both the
theory and the result of a more recent measurement \cite{bushaw:03}.
The comparison of theoretical calculations with the high-precision
experiment for the ground state of Be$^+$ \cite{wineland:83:prl} is
complicated by the existence of two different values for the nuclear
magnetic moment \cite{stone:05}. The smaller value yields a better
agreement with our theoretical result, but there is still a $2\sigma$
deviation present. Having in mind that the experimental results for
the magnetic moment are in significant disagreement with each other,
one can surmise the presence of underestimated systematic effects in
one or both of these measurements. We thus employ the comparison
presented in Table~\ref{tab:results} to infer an independent value of
the magnetic moment, which reads $\mu(^9{\rm Be})/\mu_N =
-1.177\,30(6)$ and is somewhat smaller than the both values from
\cite{stone:05}.

In summary, we have performed a large-scale CI calculation of the hfs splitting of
the $2\,^2S$ and $3\,^2S$ states of Li and Be$^+$. The results obtained from the
Dirac-Coulomb-Breit Hamiltonian agree with the previously reported nonrelativistic
values but are more accurate due to a rigorous treatment of the relativistic
correction. The QED, nuclear magnetization distribution, recoil, and
negative-continuum corrections were evaluated separately and added to the
Dirac-Coulomb-Breit value. Detailed comparison with the earlier calculations were
made and some inconsistencies in their previous treatment of individual corrections
were revealed. The calculational results for Li are in good agreement with the
experimental data. For Be$^+$, the theoretical prediction deviates from the
experimental value by 2 or $3\sigma$, depending on the value of the nuclear magnetic
moment used.

Krzysztof Pachucki is gratefully acknowledged for suggesting the topic of this
investigation and valuable discussions. The work was supported by the RFBR grant
no.~06-02-04007 and by the foundation ``Dynasty''.


\end{document}